# Unconventional Weyl exceptional contours in non-Hermitian photonic continua

QINGHUI YAN,[1,2] QIAOLU CHEN[1,2], LI ZHANG[1,2], RUI XI[1,2], HONGSHENG CHEN[1,2,3], YIHAO YANG[1,2,4]

[1] *Interdisciplinary Center for Quantum Information, State Key Laboratory of Modern Optical Instrumentation, ZJU-Hangzhou Global Scientific and Technological Innovation Center, Zhejiang University, Hangzhou 310027, China.*
[2] *International Joint Innovation Center, Key Lab. of Advanced Micro/Nano Electronic Devices & Smart Systems of Zhejiang, The Electromagnetics Academy at Zhejiang University, Zhejiang University, Haining 314400, China.*
[3] *hansomchen@zju.edu.cn*
[4] *yangyihao@zju.edu.cn*

**Abstract:** Unconventional Weyl points with topological charges higher than 1 can transform into various complex unconventional Weyl exceptional contours under non-Hermitian perturbations. However, theoretical studies of these exceptional contours have been limited to tight-binding models. Here, we propose to realize unconventional Weyl exceptional contours in photonic continua—non-Hermitian anisotropic chiral plasma, based on ab initio calculation by Maxwell's equations. By perturbing in-plane permittivity, an unconventional Weyl point can transform into a quadratic Weyl exceptional circle, a Type-I Weyl exceptional chain with one chain point, a Type-II Weyl exceptional chain with two chain points, or other forms. Realistic metamaterials with effective constitutive parameters are proposed to implement these unconventional Weyl exceptional contours. Our work paves a way toward exploration of exotic physics of unconventional Weyl exceptional contours in non-Hermitian topological photonic continua.

## 1. Introduction

Photonic Weyl point (WP) [1-14] is the linear point crossing of two bands in three-dimensional (3D) momentum space. As the sources or drains of the Berry flux with $C = +1$ (or $C = -1$, where $C$ is the Chern number) topological charges, WPs are characterized by the helicoid surface states and the discontinuous Fermi arcs at surface boundaries of photonic media [1-4, 6, 7]. Later studies have shown the charge-1 WP belongs to a big family; others include but not limited to quadratic Weyl points (QWPs, $C = 2$) [8, 15-19], spin-1 WPs ($C = 2$) [17, 20-22], 3D Dirac points ($C = 2$) [23-25], triple WPs ($C = 3$) [8, 26], and quadruple WPs ($C = 4$) [27, 28]. Members with $C > 1$ charges in the family are generally dubbed the unconventional WPs.

The photonic WPs (conventional and unconventional) can be realized in periodic, precisely-engineered, artificial structures, such as photonic crystals [5] and optical waveguide arrays [10], or in photonic continua, such as magnetized semiconductors [13] and metamaterials [12]. In comparison to the periodic structures, the electrodynamics of continua is much simpler, which significantly facilitates to gain deeper physical insights of the topological photonics. Besides, many interesting phenomena associated with the photonic

Weyl media have been observed, including chiral zero modes [29], robust surface states [8, 9], and topological self-collimations [19].

The recent rapid development of the non-Hermitian topological band theory [30-32] has brought researchers' attentions to the non-Hermitian generalizations of photonic WPs. For example, under non-Hermitian perturbations, a conventional WP can transform into a Weyl exceptional ring (WER), along which the eigenmodes of two degenerate bands coalesce into one, forming a ring of exceptional points [31]. Such WERs are proposed to exist in periodic structures decorated with gain or loss (e.g., non-Hermitian photonic crystals [33] and non-Hermitian waveguide arrays [34]), and non-Hermitian continua (e.g., lossy magnetized plasma) [35, 36]. Interestingly, the WER preserves the topological charge as well as the surface-wave Fermi arcs [32, 35]. Besides, recent theoretical works have suggested the unconventional WP under non-Hermitian perturbations can transform into various complex one-dimensional closed exceptional contours, with the topological charges preserved [32]. Several tight-binding models have been proposed to implement those unconventional exceptional contours [32, 37]. However, in realistic photonic media, such as photonic crystals and metamaterials, the modes are usually not tightly bound to any site; the band structures of photonic crystals are due to the multiple Bragg scattering in most cases [38]. Therefore, it is important to establish a framework to study the unconventional Weyl exceptional contours beyond the tight-binding models, especially in the context of photonics.

Here, we study various unconventional Weyl exceptional contours in photonic continua— non-Hermitian chiral plasma, based on the first-principle Maxwell's equations. Interestingly, by considering various perturbations to the in-plane permittivity, including non-Hermitian biaxial perturbations, Hermitian biaxial perturbations, and non-Hermitian non-reciprocal perturbations, we obtain a 3D mapping to sketch all the Hermitian/non-Hermitian generalization of an unconventional WP, more specifically, a QWP. As shown in Fig. 1, we systematically identify various forms of unconventional Weyl exceptional contours, including the quadratic Weyl exceptional rings (QWERs), two separated WERs, and the Weyl exceptional chains (WECs), denoted by the red and grey regions and the critical boundary in between, respectively. Among them, we focus on three special cases that are pinned in certain planes by the pseudo-Parity-Time (pseudo-$PT$) symmetry, which are in-plane QWERs, Type-I WECs with one chain point, and Type-II WECs with two chain points, respectively, as underlined in Fig. 1. To implement the above three cases, we then propose realistic metamaterials with effective constitutive parameters.

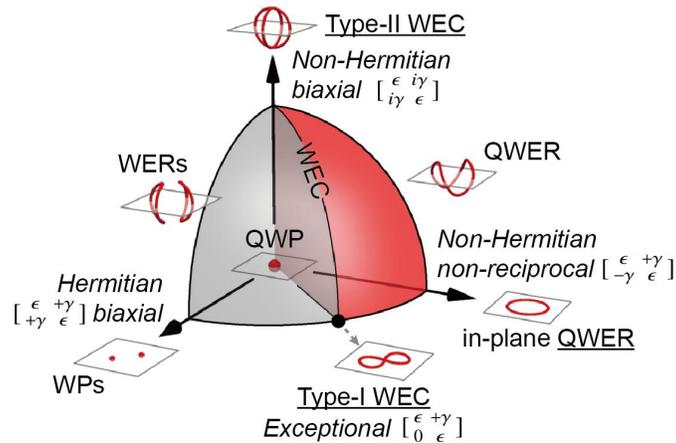

**Fig. 1.** Unconventional Weyl exceptional contours in non-Hermitian anisotropic chiral plasma, parameterized by the in-plane permittivity. By whether the contour splits or not, the space is divided into the red part, the grey part, and the boundary in between, corresponding to the

QWERs, two separated WERs, and the WECs. Underlined are three special cases that pinned in planes by pseudo-*PT* symmetries: in-plane QWERs, Type-I WECs with a single chain point, and Type-II WECs with two chain points.

## 2. Quadratic Weyl point in Hermitian anisotropic chiral plasma

To realize unconventional Weyl exceptional contours, we start by constructing a QWP, which is at the origin of the 3D parameter space (see Fig. 1). We consider a piece of plasma continua with the in-plane (in the *xOy* plane) conductivity described by the lossless Drude model $\sigma = i\omega_p^2/\omega$ ($\omega_p$ is the plasma frequency), and consider chirality [39] to break all mirror symmetries [40]. Then, using the auxiliary field method [41-43], we have

$$\begin{bmatrix} i\nabla \times & -i\mathbf{g} & \\ -i\nabla \times & & \\ i\mathbf{g}^T & & \end{bmatrix} \begin{bmatrix} \mathbf{E} \\ \mathbf{H} \\ \mathbf{J} \end{bmatrix} = \omega \begin{bmatrix} \varepsilon & +i\chi & \\ -i\chi & \mu & \\ & & \omega_p^{-2} \end{bmatrix} \begin{bmatrix} \mathbf{E} \\ \mathbf{H} \\ \mathbf{J} \end{bmatrix} \quad (1)$$

where $\varepsilon$, $\mu$, and $\chi$ are permittivity, permeability, and chirality tensors, respectively. **E**, **H**, and **J** are electric field, magnetic field, and current density, respectively. **g** is a matrix denoting in-plane conductivity, i.e., **g** = [1, 0; 0, 1; 0, 0].

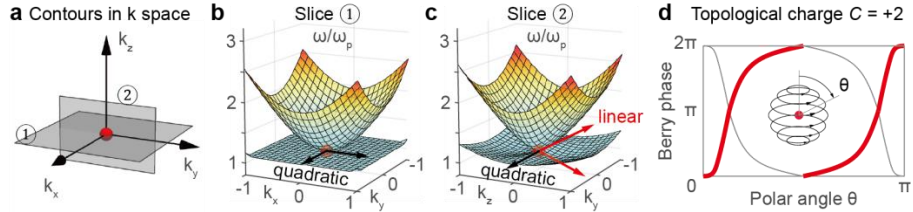

**Fig. 2.** Quadratic Weyl point in Hermitian anisotropic chiral plasma. (a) Isofrequency contour at the plasma frequency in the *k* space, where QWP is at the *k* = 0 point. (b) Quadratic in-plane dispersion. (c) Linear out-of-plane dispersion. (d) As $\chi < 0$, the accumulated Berry phase of the circle over the sphere that encloses QWP is $+4\pi$ for the upper band (in red), indicating $C = +2$.

From Eq. (1) we can qualitatively calculate the photonic bands of the continua. Without losing the generality, we set $\varepsilon = \mu = \omega_p = 1$ and $\chi = -0.4$, and $\nabla \sim i\mathbf{k}$ for the continua. As a result in Fig. 2(a), a QWP exists at the *k* = 0 point at the modified plasma frequency $\omega'_p = \omega_p/(1 - \chi^2)^{0.5}$. As it named, the in-plane dispersion is quadratic, indicating the photon is effectively massive due to the sufficient light-matter interaction with the electronic gas. The calculation of topological charge also proves the QWP: we enclose the *k* = 0 point by a small sphere in momentum space and calculate Wilson loops for different polar angles of the sphere. As shown in Fig. 2(b), as the angle varies from 0 to $\pi$, the accumulated phase is $+4\pi$ for the upper band of the point degeneracy (in red), indicating $C = +2$ for $\chi < 0$ (or $C = -2$ for $\chi > 0$), which is consistent with our understanding of the QWP.

## 3. Unconventional Weyl exceptional contours in non-Hermitian anisotropic chiral plasma

With the Hermitian QWP on hand, we still need a strategy to break the Hermiticity for the Weyl exceptional contours. Motivated by Ref. [44], we use the perturbative method to expand the band structure of QWP in the local momentum space and see the variation of eigenfrequencies with the constitutive parameters. We rewrite Eq. (1) as

$$-\left(\mathbf{k}\times+i\chi\omega\right)\mu^{-1}\left(\mathbf{k}\times+i\chi\omega\right)\mathbf{E}+\omega_p^2\mathbf{g}\mathbf{g}^T\mathbf{E}=\omega^2\varepsilon\mathbf{E}, \tag{2}$$

and set $\mu = 1$ for simplicity. Due to the in-plane longitudinality of the wavefunctions at the QWP ($E_z = 0$), we neglect this component in Eq. (2). By perturbing both the wavevector and the permittivity tensor, the local Hamiltonian around the plasma frequency is (see more detail in Appendix)

$$H = H_{\text{QWP}} - \delta\varepsilon_\parallel, \tag{3}$$

where $\delta\varepsilon_\parallel$ is the (zeroth-order) perturbation of the in-plane permittivity under Cartesian coordinates, and $H_{\text{QWP}}$ is the local Hamiltonian of QWP, i.e.,

$$H_{\text{QWP}} = \left(k_y^2 - k_x^2\right)\sigma_3 - 2k_xk_y\sigma_1 - \text{sgn}(\chi)k_z\sigma_2, \tag{4}$$

where $\sigma_n$, $n = 1, 2, 3$ are the Pauli matrices, and sgn($\chi$) gives the sign of chirality.

Equation (3) bridges the local Hamiltonian and the constitutive parameters of the continua; by tuning the in-plane permittivity, we perturb the local Hamiltonian to turn the QWP into various quadratic Weyl exceptional contours. The perturbation term can be decomposed by the four Pauli matrices with complex coefficients, denoted as $\gamma_n\sigma_n$, where $n = 0, 1, 2, 3$, and $\gamma_n$ is complex. Therefore, there are eight degrees of freedom in total, which we are going to discuss respectively in the following text.

*3.1 Hermitian perturbations to the in-plane permittivity*

To analyze the Hermitian perturbations to the QWP, we refer to the idea of topological photonics, which treats QWP as the superposition of two WPs pinned by the rotational symmetry [26, 45]. Since the WP is robust against any Hermitian perturbation, there are no other ways of evolution than to split the QWP into two WPs. More specifically, the $\gamma\sigma_0$ term ($\gamma$ is a positive real number) performs a trivial perturbation that simply changes the frequency of QWP. The $\gamma\sigma_2$ term shifts the QWP along the $k_z$ direction, corresponding to the gyrotropic material with a magnetic axis in $z$ direction that breaks the time-reversal ($T$) symmetry. As for the $\gamma\sigma_1$ term (or $\gamma\sigma_3$ term, which is equivalent by $\pi/4$ rotation along the $z$ axis), which is biaxial anisotropic that breaks the continuous rotational symmetry along $z$ direction ($C_{\infty z}$) down to two-fold rotational symmetry $C_{2z}$, the QWP splits into two in-plane WPs, as shown in Fig. 3(a).

*3.2 Non-Hermitian perturbations to the in-plane permittivity*

It has been known that the WER is the non-Hermitian generalization of a WP [31], and the QWP is the superposition of two identical WPs [26], so the non-Hermitian generalization of a QWP should be two WERs in essence. In this sense, we consider the quadratic Weyl exceptional contour as the connection and merging of the two WERs.

We start the analysis with the $i\gamma\sigma_0$ term that corresponds to the addition of an isotropic lossy term onto the permittivity. The lossy term keeps the degeneracy of the QWP, even though the eigenfrequency is complex. This can be understood from the lossy Drude model, which does not lead to exceptional contours at the plasma frequency.

For the $i\gamma\sigma_2$ term, as shown in Fig. 3(b), the two WPs inside the QWP expand into two identical WERs to form a QWER. Moreover, we find the $C_{2z}T$ symmetry acts as a pseudo-$PT$ symmetry in the $k_z = 0$ plane. Given a mode labelled with ($k_x$, $k_y$, $k_z$, $\omega$), the $C_{2z}$ symmetry maps the mode into another mode labelled with (-$k_x$, -$k_y$, $k_z$, $\omega$), while the $T$ symmetry into (-$k_x^*$, -$k_y^*$, -$k_z^*$, $\omega^*$). Thus, at the $k_z = 0$ plane, $C_{2z}T$ together maps ($k_x$, $k_y$, 0, $\omega$) to ($k_x$, $k_y$, 0, $\omega^*$), which is identical to the *PT* symmetry (we thus refer to it as the pseudo-*PT* symmetry hereafter). Therefore, modes in the in-plane part of the band are either in an exact-pseudo-*PT* phase where eigenfrequencies are all real numbers, or in a broken-pseudo-*PT* phase where bands are coupled in pairs with conjugate eigenfrequencies; the QWER lies at the boundary between two phases. As shown in Fig. 3(b), the red circle is the exceptional ring, and the red area is of broken-pseudo-*PT* phase.

For the $i\gamma\sigma_1$ term (or the $i\gamma\sigma_3$ term), as shown in Fig. 3(c), the QWP transforms into two orthogonal WERs which cross with each other at two points along the $k_z$ axis. We thus refer to it as the Type-II WEC. *T* symmetry no longer exists for this material; instead are the $C_{2x}T$ and the $C_{2y}T$ symmetries acting as pseudo-*PT* symmetries that pin the two WERs in the $k_x = 0$ plane and the $k_y = 0$ plane, respectively. Moreover, two WERs are related by $C_{4z}T$, so we only plot one of them in Fig. 3(c).

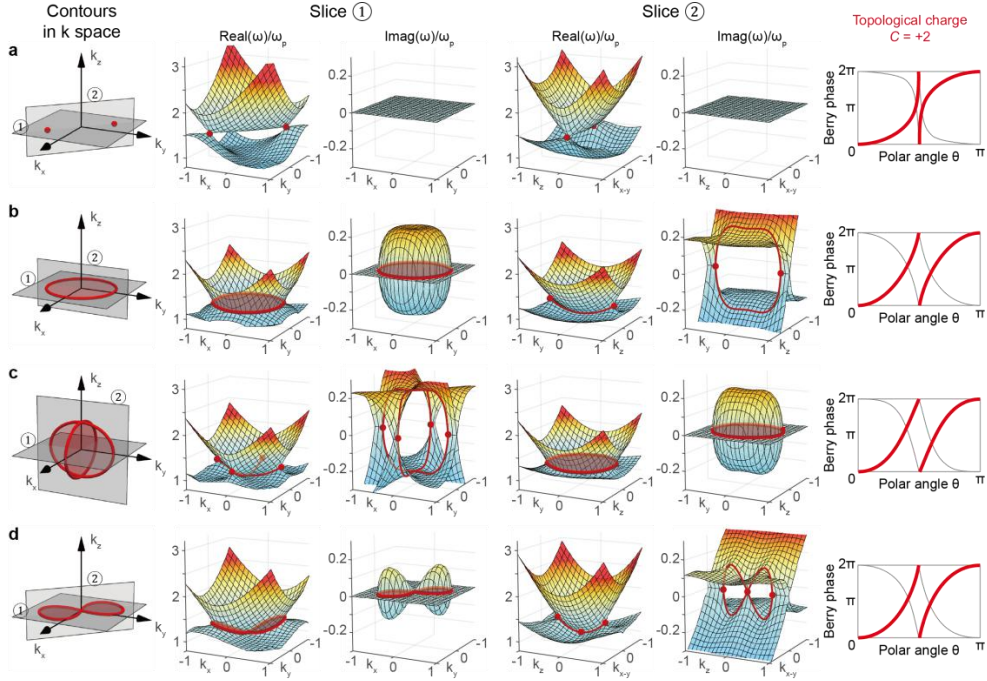

**Fig. 3.** Four typical cases evolving from the QWP. (a) Two WPs, where $\varepsilon_\parallel = [1, 0.4; 0.4, 1]$. (b) QWER, where $\varepsilon_\parallel = [1, 0.4; -0.4, 1]$. (c) Type-II WEC, where $\varepsilon_\parallel = [1, 0.4i; 0.4i, 1]$. (d) Type-I WEC, where $\varepsilon_\parallel = [1, 0.4; 0, 1]$. For each case we plot the exceptional contour, the band diagrams in the slices indicated in the left panel, and the Wilson loop on a sphere covering the unconventional Weyl exceptional contours.

### *3.3 Unconventional Weyl exceptional contours in the in-plane permittivity parameter space*

Above, we have discussed all eight types of perturbations, respectively. Only three of them leads to nontrivial deformations, i.e., Real($\gamma_1$), Imag($\gamma_2$), and Imag($\gamma_1$), where Real(·) and Imag(·) give the real and imaginary parts of the input, respectively, as shown in Tab. 1. The rest perturbations merely shift, rotate, or tilt the contour that does not change the connection. Therefore, we can study the unconventional Weyl exceptional contours in a 3D parameter space corresponding to the in-plane permittivity tensor. Recalling Fig. 1, besides the exceptional contours along the three axes and the original point, we calculate the exceptional contours for the rest part of the space, and pick out three intermediate states between every two axes to depict how the exceptional contour evolves as the in-plane permittivity varies. Depending on whether the QWER splits into two WERs or not (Real($\gamma_1$) - Imag($\gamma_2$) >0 or <0), the space can be divided into two regions (grey and red). At the critical boundary (Real($\gamma_1$) - Imag($\gamma_2$) = 0), two WERs are critically connected to form the WEC.

Besides the Type-II WEC, we find another interesting case on the boundary, namely the Type-I WEC, where the two WERs are in the same plane (pinned by $C_{2z}T$) with only one chain point at the $k = 0$ point (pinned by $C_{2z}$), as shown in Fig. 3(d) and the fifth column in Tab. 1. Due to the same strength of perturbation by $\sigma_1$ and $i\sigma_2$, the perturbation term is upper-triangular, which is the sign of exceptional structures. Consequently, the in-plane permittivity is exceptional with coalesced eigenmodes at the chain point at the $k = 0$ point. In a broader sense, the chain point is pinned at the $k = 0$ point if and only if

$$\gamma_1^2 + \gamma_2^2 + \gamma_3^2 = 0 \tag{5}$$

where $\gamma_{1,2,3}$ are complex numbers that should not be zero simultaneously, otherwise the perturbation term vanishes.

| Contours | Two WPs | In-plane QWER | Type-II WEC | Type-I WEC |
|---|---|---|---|---|
| Figures | Fig. 3(a) | Fig. 3(b) | Fig. 3(c) | Fig. 3(d) |
| Perturbation | $\gamma\sigma_1$ | $i\gamma\sigma_2$ | $i\gamma\sigma_1$ | $\gamma(\sigma_1+i\sigma_2)$ |
| $\varepsilon_\parallel =$ | $\begin{bmatrix} \varepsilon & +\gamma \\ +\gamma & \varepsilon \end{bmatrix}$ Hermitian biaxial | $\begin{bmatrix} \varepsilon & +\gamma \\ -\gamma & \varepsilon \end{bmatrix}$ Non-Hermitian non-reciprocal | $\begin{bmatrix} \varepsilon & +i\gamma \\ +i\gamma & \varepsilon \end{bmatrix}$ Non-Hermitian biaxial | $\begin{bmatrix} \varepsilon & +2\gamma \\ 0 & \varepsilon \end{bmatrix}$ Non-Hermitian exceptional |
| Symmetries | $T$, $C_{2z}$, $\underline{C_{2z}T}$ | $T$, $C_{\infty z}$, $\underline{C_{2z}T}$ | $C_{4z}T$, $\underline{C_{2x}T}$, $\underline{C_{2y}T}$ | $T$, $C_{2z}$, $\underline{C_{2z}T}$ |

**Table 1.** Variants of QWP when introducing perturbations to the in-plane permittivity, where $\varepsilon$ and $\gamma$ are real numbers. Underlined are the pseudo-PT operators that respectively pin the exceptional contours in one of the $k_\alpha = 0$ planes, where $\alpha = x, y, z$.

## 4. Metamaterial design to achieve the unconventional Weyl exceptional contours

In the following, we design a metamaterial structure to implement the QWP, and then add perturbations to the background permittivity to achieve the unconventional Weyl exceptional contours. As shown in Fig. 4(a), the unit cell is cuboid with periods $p_x = p_y = 7$ *mm*, $p_z = 10$ *mm* along *x*, *y*, and *z* directions, respectively. In each unit cell, there is a metallic resonant structure made with 0.7 *mm* thickness rods. The structure can be considered as the deformation of two in-plane split-ring resonators, described by the Drude-Lorentz model with a resonant frequency at about 6.3 GHz. Besides, the slits of the split ring resonators are offset from the center point of the skeleton. In doing so, all mirror symmetries are broken and the chirality is introduced. Here, the structural parameters are $d = 0.7$ *mm*, $h_x = h_y = 6.0$ *mm*, $h_z = 2.9$ *mm*, and *offset* = 1.0 *mm*. The spatial group of the unit cell is No. 89 (*P422*). Suppose the background material has $\varepsilon = 4$ and $\mu = 1$, then at the $\Gamma$ point ($k = 0$), the third and the fourth bands intersect as a QWP. Note that the QWP has been experimentally realized at microwave [19] and infrared frequencies [15].

Then, by tuning the in-plane permittivity of the background material, different exceptional contours emerge as predicted. As shown in Fig. 4(b), $\varepsilon_{xx} = \varepsilon_{yy} = 4$ and $\varepsilon_{xy} = -\varepsilon_{yx} = 0.5$, i.e., the

$i\gamma_2\sigma_2$ type, and thus the QWP expands into a QWER. Due to the $C_4$ symmetry, the contour is not a perfect circle. By breaking the $C_{2x}$ and $C_{2y}$ symmetries, the magnetic spatial group falls down to No. 75.2 (*P41'*).

As shown in Fig. 4(c), if we apply $\varepsilon_{xy} = \varepsilon_{yx} = 0.5i$ to the background media, i.e., the $i\gamma_1\sigma_1$ type, the QWP turns into a Type-II WEC. Due to the strong anisotropy, the chain is greatly pressed along $z$ direction. The magnetic spatial group is No. 89.91 (*P4'2'2*).

As shown in Fig. 4(d), as $\varepsilon_{xy} = 0.5$ and $\varepsilon_{yx} \approx -0.06$, the QWP turns into a Type-I WEC that pinned by the $C_{2z}T$ symmetry with chain point at the $k = 0$ point. The in-plane permittivity is slightly biased from upper-triangular, because $k$ is comparable to $2\pi/p_x$ or $2\pi/p_y$ (the length of the Brillouin zone), the non-local effect more or less induces higher-ordered dispersions that twist the photonic bands. The effect is negligible as the size of the Brillouin zone is finite. The magnetic spatial group is No. 3.2 (*P21'*).

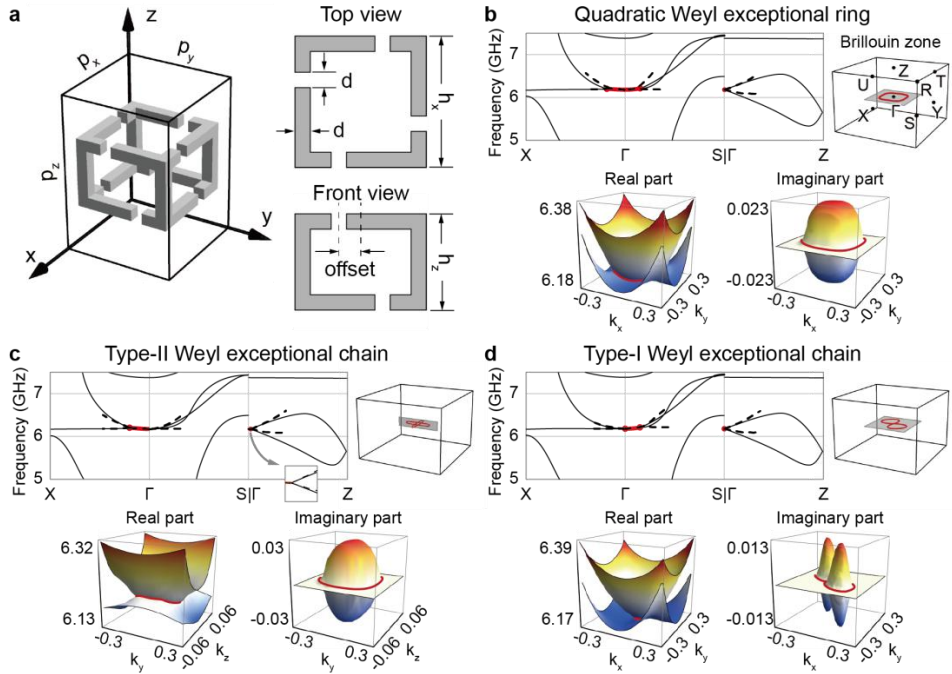

**Fig. 4.** Metamaterial designs to implement various unconventional Weyl exceptional contours. a) Unit cell of metamaterial. The metallic structure (in grey) is designed to create in-plane resonance to introduce chirality. By switching the in-plane permittivity of the background material to [4.0, 0.5; -0.5, 4.0], [4.0, 0.5$i$; 0.5$i$, 4.0], and [4.0, 0.5; -0.06, 4.0], respectively, the third and the fourth bands intersect as b) the QWER, c) the Type-II WEC, and d) the Type-I WEC around 6.3 GHz in the vicinity of the $\Gamma$ point, respectively. All exceptional contours are pinned inside one of the $k_\alpha = 0$ planes by $C_{2\alpha}T$ symmetries, where $\alpha = x, y, z$. The grey area in the Brillouin zone in (b-d) cover the exceptional contour, where a detailed band plot is given to show the real and imaginary part of the eigenfrequencies. $k_x$ and $k_y$ are normalized by $\pi/p_x$ and $\pi/p_y$, respectively. For each case in (b-d), we fit the bulk bands by the continua model with effective constitutive parameters, plotted in the cyan curves in the vicinity of the $\Gamma$ point. Note in (d), in-plane permittivity is not rigorously upper-triangular due to the nonlocal effect.

Then, we fit the bulk bands of the metamaterials by the continua model. As shown Fig. 4(b-d), in the vicinity of the exceptional contours, we plot the fitting results by dashed curves, with constitutive parameters $\varepsilon = 4$ and $\omega_p = 2\pi \times 12.2$ GHz. Due to the magnetic resonance of the split ring resonator, the permeability is greater than 1, i.e., $\mu_{xx} = \mu_{yy} = 4$, $\mu_{zz} = 7$. The chirality is $\chi = 0.7$. One can observe that different perturbations lead to different off-diagonal elements of the effective permittivity tensor. Here, we have $\varepsilon_{xy} = -\varepsilon_{yx} = -0.05$ for the QWER, $\varepsilon_{xy} = \varepsilon_{yx} = 0.04i$ for the Type-II WEC, and $\varepsilon_{yx} = -0.07$ for the Type-I WEC.

## 5. Discussion

We have thus theoretically identified various unconventional Weyl exceptional contours in the non-Hermitian photonic continua, including the QWER (the exact superposition of two WERs), the Type-I WEC with one chain point and the Type-II WEC with two chain points. Based on the Maxwell's equations and the perturbative method, we bridge the local Hamiltonian of unconventional Weyl exceptional contours with the constitutive parameters of the non-Hermitian photonic media. Several metamaterials with effective constitutive parameters have been designed to implement the unconventional Weyl exceptional contours. Besides, it would be interesting to investigate the directional amplification based on the QWER with broken Hermiticity and reciprocity while preserving the $T$ symmetry [46], and the arbitrary gain/loss control of light polarization based on the WEC with complex symmetric anisotropic permittivity [47]. With the recent rapid development in photonic metamaterials [48-50], it is possible to experimentally implement those unconventional Weyl exceptional contours in the future. Finally, our work sheds light on the exploration of exotic physics of unconventional Weyl exceptional contours in non-Hermitian topological photonic continua, such as non-Hermitian skin effect [51], and non-Hermitian surface-wave Fermi arcs.

**Funding.** The work at Zhejiang University was sponsored by the National Natural Science Foundation of China (NNSFC) under Grants No. 61625502, No. 11961141010, No. 61975176, and No. U19A2054, the Top-Notch Young Talents Program of China, and the Fundamental Research Funds for the Central Universities.

**Supplemental document.** See Appendix for supporting content.

**Disclosures.** The authors declare no conflicts of interest.

**Data Availability.** Data underlying the results presented in this paper are not publicly available at this time but may be obtained from the authors upon reasonable request.

## APPENDIX: Intuitive derivation on the local Hamiltonian of the quadratic Weyl point and the perturbation by the permittivity

The perturbed local Hamiltonian provides a brief formula to estimate the variation of photonic bands caused by the variation of constitutive parameters [44]. In our case, the local Hamiltonian is an operator whose eigenvalues are the small variations of angular frequencies in the vicinity of $\omega'_p$ at the $k = 0$ point. To derive the local Hamiltonian of the quadratic Weyl point, we first write down Maxwell's equations with the auxiliary field.

$$\begin{bmatrix} & +i\nabla\times & -i\omega_p \mathbf{g} \\ -i\nabla\times & & \\ +i\omega_p \mathbf{g}^T & & \end{bmatrix} \begin{bmatrix} \mathbf{E} \\ \mathbf{H} \\ \mathbf{J} \end{bmatrix} = \omega \begin{bmatrix} \varepsilon & +i\chi & \\ -i\chi & \mu & \\ & & 1 \end{bmatrix} \begin{bmatrix} \mathbf{E} \\ \mathbf{H} \\ \mathbf{J} \end{bmatrix} \qquad (7)$$

In a homogeneous material, $\nabla \sim i\mathbf{k}$. Suppose $\varepsilon_0 = \mu_0 = 1$ and $\mu = 1$, and move the chirality term to the left side

$$\begin{bmatrix} & -\mathbf{k}\times -i\chi\omega & -i\omega_p\mathbf{g} \\ +\mathbf{k}\times +i\chi\omega & & \\ +i\omega_p\mathbf{g}^\dagger & & \end{bmatrix}\begin{bmatrix}\mathbf{E}\\\mathbf{H}\\\mathbf{J}\end{bmatrix} = \omega\begin{bmatrix}\varepsilon & & \\ & 1 & \\ & & 1\end{bmatrix}\begin{bmatrix}\mathbf{E}\\\mathbf{H}\\\mathbf{J}\end{bmatrix} \quad (8)$$

Rewrite Eq. (8) in the form of $D\psi = \omega M\psi$. Then doubling the operator $M^{-1}D$ for $M^{-1}DM^{-1}D\,\psi = \omega^2\psi$ to decouple the electric field and the magnetic field. Pick out the electric part, we have

$$-(\mathbf{k}\times +i\chi\omega)^2\mathbf{E} + \omega_p^2\mathbf{gg}^T\mathbf{E} = \omega^2\varepsilon\mathbf{E} \quad (9)$$

As $\varepsilon = 1$, it is Eq. (2) in the main text.

If $\varepsilon_\parallel = 1$, then at the $k = 0$ point around $\omega_p' = \omega_p / (1 - \chi^2)^{0.5}$ has a quadratic Weyl point, whose electric field has no out-of-plane component, so we cancel $E_z$.

$$\left(k_z^2 + \begin{bmatrix}k_y^2 & k_xk_y \\ k_xk_y & k_x^2\end{bmatrix} - 2\chi k_z\omega\begin{bmatrix} & -i \\ +i & \end{bmatrix}\right)\mathbf{E}_\parallel = \left(\omega^2\varepsilon_\parallel - \chi^2\omega^2 - \omega_p^2\right)\mathbf{E}_\parallel \quad (10)$$

where the subscript $\parallel$ denotes the field and the tensor with it are in-plane.

To derive the local Hamiltonian, suppose a small variation of angular frequency $\delta\omega$ from $\omega_p'$, we have

$$\left(k_z^2 + \begin{bmatrix}k_y^2 & k_xk_y \\ k_xk_y & k_x^2\end{bmatrix} - 2\chi k_z\omega'_p\begin{bmatrix} & -i \\ +i & \end{bmatrix}\right)\mathbf{E}_\parallel$$
$$= \left(2(1-\chi^2)\omega'_p\,\delta\omega + \underline{2\chi k_z\delta\omega\begin{bmatrix} & -i \\ +i & \end{bmatrix}}\right)\mathbf{E}_\parallel \quad (11)$$

In the vicinity of the $k = 0$ point, the underlined term is small. By neglecting this term and other coefficients that do not affect the topological properties, we have the local Hamiltonian of the quadratic Weyl point.

$$H_{\mathrm{QWP}} \sim k_z^2 + \begin{bmatrix}k_y^2 & -k_xk_y \\ -k_xk_y & k_x^2\end{bmatrix} - \frac{1}{2}\mathrm{sgn}(\chi)k_z\begin{bmatrix} & -i \\ +i & \end{bmatrix}$$
$$= \underline{\left(k_z^2 + \frac{1}{2}k_x^2 + \frac{1}{2}k_y^2\right)\sigma_0} + \frac{1}{2}(k_y^2 - k_x^2)\sigma_3 - k_xk_y\sigma_1 - \frac{1}{2}\mathrm{sgn}(\chi)k_z\sigma_2 \quad (12)$$

which is Eq. (4) in the main text.

As the perturbation term is added to the permittivity, the variation of $\omega^2\varepsilon_\parallel$ on the right side of Eq. (10) is $\delta(\omega^2\varepsilon_\parallel) = 2\omega_p'\delta\omega\varepsilon_\parallel + \underline{\omega_p'^2\delta\varepsilon_\parallel}$, where is underlined term is nonzero. Following the same procedure as above and neglecting some coefficients, we have

$$H_{\mathrm{QWP}}\mathbf{E} = \delta\omega\mathbf{E} + \delta\varepsilon_\parallel\mathbf{E} \quad (13)$$

which leads to Eq. (3) in the main text.

Our intuitive derivation is equivalent to the rigorous degenerate state perturbative method by auxiliary field method with Maxwell's equations [52].